# Singlet Fission Photovoltaics: Progress and Promising Pathways

Alexander J. Baldacchino[1], Miles I. Collins[2], Michael P. Nielsen[1], Timothy W. Schmidt[3], Dane R. McCamey[2], Murad J. Y. Tayebjee[1]

[1]School of Photovoltaic and Renewable Energy Engineering, UNSW, Sydney, New South Wales, Australia

[2]ARC Centre of Excellence in Exciton Science, School of Physics, UNSW, Sydney, New South Wales, Australia

[3]ARC Centre of Excellence in Exciton Science, School of Chemistry, UNSW, Sydney, New South Wales, Australia

# ABSTRACT

Singlet fission is a form of multiple exciton generation which occurs in organic chromophores when a high energy singlet exciton separates into two lower energy triplet excitons, each with approximately half the singlet energy. Since this process is spin-allowed it can proceed on an ultrafast timescale of less than several picoseconds, outcompeting most other loss mechanisms and reaching quantitative yields approaching 200%.

Due to this high quantum efficiency, the singlet fission process shows promise as a means of reducing thermalisation losses in photovoltaic cells. This would potentially allow for efficiency improvements beyond the thermodynamic limit in a single junction cell. Efforts to incorporate this process into solar photovoltaic cells have spanned a wide range of device structures over the past decade. In this review we compare and categorise these attempts in order to assess the state of the field and identify the most promising avenues of future research and development.

# 1. INTRODUCTION
## 1.1 Early Singlet Fission Studies

Singlet fission (SF) is a form of multiple exciton generation wherein an optically prepared singlet exciton splits into two triplet excitons on neighbouring chromophores:

$$S_0 + S_1 \rightarrow (TT) \rightarrow T_1 + T_1 \quad (1)$$

$$S_0 + S_1 \rightarrow T_1 + T_1 \quad (2)$$

where $S_0$ is the ground state chromophore, $S_1$ is the lowest energy singlet excited state of the chromophore, TT is a coupled triplet pair and $T_1$ is a chromophore in the lowest energy triplet excited state.

This process occurs in organic semiconductors that meet the energetic criterion:

$$E_{S1} \geq 2E_{T1} + nk_BT \quad (3)$$

where $E_{S1}$ and $E_{T1}$ are the energies of the first excited singlet and triplet excitons, respectively. The second term accounts for endothermic fission which can occur for small values of $n$ (where $k_B$ and $T$ are Boltzmann's constant and the lattice temperature).

The first period of SF research was focussed on spectroscopic studies. The first experimental observation of this process was by Singh *et al.* in 1965, through delayed fluorescence measurements in anthracene crystals[1]. Comparison between the fluorescence detected from the anthracene crystals under laser excitation indicated a doubling of triplet generation when switching between the first and second harmonics of the laser source, at 694nm and 347 nm respectively. Only the latter of these excitation wavelengths was sufficient in energy to induce direct one-photon absorption to the singlet excited state in anthracene. The interpretation of the result therefore was that singlet excitons in anthracene were undergoing SF to the triplet excited state. Studies of SF systems continued into the late 1970's, confirming the presence of SF in tetracene[2–5] and perylenes[5]. Research during this period consisted of spectroscopic investigations of the phenomenon, including fluorescence, magnetic field effect, and scintillation experiments. One of the most significant works during this period was the investigation by Merrifield *et al.* in 1971[6], which characterised the effect of a magnetic field on the prompt and delayed fluorescence in tetracene. The description of spin dependent processes in this system have formed the basis for the magnetic field dependent spectroscopic techniques discussed in Section 3.1.2.

In the early 1980s, SF was demonstrated to occur in some organic systems such as photosynthetic bacteria[7,8], as will be further discussed in Section 2.1.4. Aside from this, however, SF gradually faded out of the literature for many years. This occurred despite Dexter's 1979 suggestion that SF could be employed to augment the current of photovoltaic cells[9], likely due to the relative infancy of PV technology at the time. This review is concerned with the second period of SF research: its application to solar energy harvesting.

## 1.2 The Thermodynamics of Solar Energy Generation
### 1.2.1 Unavoidable Losses from Single Threshold Devices

In the early 2000's, as solar energy technology began to mature and be deployed commercially, researchers (particularly chemical physicists or physical chemists) drove a resurgence of interest in SF. Using Dexter's initial proposition as a motivator, research into both the thermodynamic potential of SF-augmented solar cells and the photophysics of SF accelerated.

To understand the former, we must first consider the fundamental thermodynamic limitations of conventional singlet threshold photovoltaic devices using detailed balance. This limit is dependent on the band gap, $E_g$, and was calculated to be 30% for a band gap of 1.1 eV by Queisser *et al.* under 6000 K blackbody radiation[10–12]. In 2011 Hirst and Ekins-Daukes elegantly classified the unavoidable loss mechanisms into five loss processes, as shown in Figure 1 [11]. These are briefly described below:

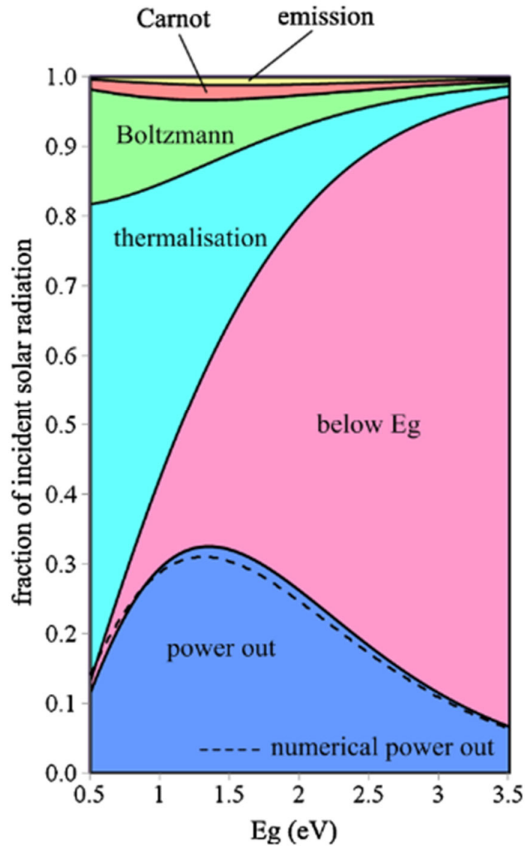

Figure 1: Fractions of incident solar radiation, extractable as electric power and losses to thermodynamic processes as a function of cell band gap energy. Diagram from Hirst *et al.*[11], Reproduced with permission from Wiley Materials, Copyright (2021).

Emission Losses

As a result of Kirchoff's law, absorbers of blackbody radiation must also emit radiation[10–12]. Radiative recombination of excited state electrons and holes limits the available photocurrent which can be collected by reducing carrier populations. Emission losses are most significant within the 1-2 eV range, tailing off significantly on either side.

Carnot Factor

A thermodynamic treatment of the PV cell allows it to be considered as a heat engine[12] in which the sun is a reservoir with $T_S$ and the cell is a thermal reservoir at $T_A$[11]. As with any heat engine the 2nd law of thermodynamics requires there to be some thermal transfer between the two reservoirs. This entropic loss of free energy is referred to as the Carnot losses. Carnot losses are most significant for low band gaps and begin tailing off past 1.5 eV.

Combined losses from the Carnot factor and emission do not exceed 5% of incident solar power for any reasonable PV cell band gap for solar energy harvesting. As a result of this, efforts to reduce these losses will not yield a significant change to solar PV efficiency.

Boltzmann Factor

The Boltzmann factor losses arise from the mismatch between the absorption and emission angles of the cell. Expansion of photon modes therefore results in an entropy generation process[11]. Boltzmann losses decrease as the band gap of the cell increases and account for over 10% of losses in the range of commercial PV cells.

Below $E_g$ Losses

Photons with energy $E_\gamma < E_g$ lack the energy required to photoexcite electrons out of their ground state configuration. Therefore, these photons will not generate any photocurrent in the cell. Given the broad shape of blackbody spectra, this severely limits the efficiency of photovoltaic cells[10–12]. The power lost through the failure to absorb sub-band gap photons is simply the total optical power emitted by the blackbody source in the region $E < E_g$. As a result of this, below band gap losses increase with the band gap of the PV cell. While these losses are often referred to as transmission losses, this is a misnomer; recent work has shown that silicon solar cells absorb the entire solar spectrum, but photons with $E < E_g$ contribute to heating rather than photocurrent[13].

Thermalisation losses

Photons with energy $E_\gamma > E_g$, will generate photoelectrons with kinetic energy in excess of the band gap (hot carriers). Due to strong interactions between carriers and lattice phonons, hot carriers will lose this excess energy to the absorber lattice and cool to the band-edge[11,12]. The calculated power loss from thermalisation is given by the sum of the excess energy in each photoelectron generated. The total optical power lost to thermalisation therefore increases as the band gap of the cell decreases.

### 1.2.2 The Potential Benefit of Singlet Fission Solar Cells

Since thermalisation and sub-band gap losses are negatively correlated, minimising one of these loss mechanisms through selecting a specific band energy will maximise the other, as seen in Figure 1 [11]. Design of an efficient single threshold PV cell therefore requires a compromise between thermalisation and sub-band gap losses. Even at peak efficiency, both thermalisation and sub-band gap losses account for over 50% of incident solar power. Efforts to reduce both of these processes offer the most significant promise for achieving meaningful step increases in solar PV cell efficiency.

Multiple exciton generation processes such as SF offer a means of reducing thermalisation losses when a low band gap absorber is paired with a SF chromophore with $E_{S1} \approx 2E_g$[14]. This is a result of there being two absorbing thresholds:

- In the region where $E_g < E_\gamma < E_{S1}$, photocurrent will be produced as in a typical solar cell, by direct absorption and production of a single electron hole pair by the low band gap absorber.
- In the region where $E_\gamma > E_{S1}$, the SF chromophore will absorb the incoming photons and produce two triplet excitons via SF. The photocurrent produced by absorption of photons in this region will be doubled if both triplet excitons are dissociated, effectively reducing thermalisation losses.

If used in isolation, the SF chromophore will not have an advantage over a standard chromophore of the same band gap, since the doubling of photocurrent via SF also halves the voltage of the cell.

The first calculation of the thermodynamic energy conversion efficiency limit of a multiple exciton generation solar cell was 41.9%[15], roughly equivalent to the 44.6% limit calculated for two junction series tandem cells[8]. This work was motivated by the potential of multiple exciton generation in quantum dot systems, which have only been observed to occur exothermically. If, however, one accounts for endothermic fission, the efficiency of a single junction multiple exciton generation solar cell is 45.9%[16]. This improvement can be achieved without the additional complications introduced by series tandem cells however, such as the need for current matching and tunnel junction optimisation. Semi-empirical methods using realistic rates for acenes put this limit closer to 35%[17].

The reduction in thermalisation losses afforded by SF will also contribute to a reduction in PV module temperature. Recent work by Jiang *et al*. estimated that this would lead to a sufficient decrease in thermal degradation to increase silicon cell lifetime by 3.7 years (14.9%) for a tetracene/Silicon PV cell compared to a standard silicon PV cell[18]. Given the rapid uptake of solar power generation and the urgency to switch to renewable energy sources, it is timely to review the current state of SF-augmented solar cells.

### 1.3 Scope of this review

Multiple reviews have been published over the past several years in the field of SF research. The earliest major review of the field was published in 2010 by Smith *et al*.[14], and comprehensively described spin physics, chromophores and spectroscopy in the field.

Since that period, the field has advanced and expanded significantly, leading to the publication of a number of more specialised reviews[19–25]. Multiple reviews focusing on chromophores have been written over this period, with recent reviews by Casillas *et al*.[26] and Ullrich *et al*.[22] providing an up-to-date summary of the field. For this reason, this review will not delve in depth into the development of SF chromophores.

Spectroscopic techniques[24,27], and the role of magnetic field effects on triplet pair states[20,23,28] have also been assessed in several recent review articles and so will not be the focus of this review. We will, however, provide a useful framework for assessing the efficacy of SF devices using magnetic field spectroscopy in Section 3. Finally, other forms of multiple exciton generation (i.e. in quantum dots) and optical downconverters are also outside the scope of this review.

The aim of this review will be to provide a complete overview of the state of SF photovoltaic device studies at the date of submission. The devices will be classified according to the SF acceptor used and assessed for their potential future suitability in power generation. Such an assessment of device structures in this manner has not been undertaken since the review by Rao *et al.* in 2017[25]. A book chapter by Ehrler exploring SF sensitised silicon was published as of October 2021 but did not explore all SF device architectures[29].

Given the significant recent advancements in SF device research, there is a need for a current assessment of the state of the field. The remainder of this review is set out as follows. Section 2 will outline the material requirements of SF chromophores and acceptors, as well as provide a brief overview of several chromophores of interest. Section 3 will begin by exploring methods used to verify the occurrence of SF in a photovoltaic device and quantify its contribution to device efficiency. It will then go on to discuss attempts to produce functioning SF photovoltaic devices in literature, categorised according to the acceptor architecture used. Section 4 will conclude the review with a summary of the challenges involved in realising a functional SF device and an outlook of potential future research directions.

## 2. MATERIAL REQUIREMENTS & CHROMOPHORES
### 2.1 Singlet Fission Chromophore/Acceptor Combination Requirements
#### 2.1.1 Requirements for Singlet Fission Chromophores

In the field of SF there have been significant efforts to characterise and improve the chromophores known to undergo the process. As a result of the stringent requirements for efficient SF, the number of suitable chromophores is limited.

For a SF chromophore to be practical in devices, it must possess the following qualities:

1. **High SF yield in the neat SF material** The total exciton quantum yield (QY) must be sufficiently larger than 100% for SF to provide a useful benefit to device efficiency. i.e. for each incident photon meeting the energy condition for SF, the number of excitons produced must be as close as possible to the maximum of two. SF chromophores with QY~200% are classified as undergoing quantitative SF.
2. **Fast SF rate** Related to the preceding point, but distinct when with an acceptor, the rate of SF must be sufficiently fast to outcompete the various loss processes which may be present. The most significant process is singlet exciton transfer to the acceptor. Since these decay processes will vary depending on the acceptor structure used, the minimum acceptable SF rate will also vary between device structures.
3. **Slow TTA rate** The triplet-triplet annihilation (TTA) reverse reaction must be slow relative to triplet harvesting in order SF to be used effectively in a device.
4. **Rapid (TT) state dissociation** The triplet pair state must dissociate rapidly into free triplets in order to inhibit TTA to the ground state, or excimer formation.
5. **Compatible $T_1$ & $E_g$ energies** The energy of triplet excitons produced via SF must have $E_{T1} \geq E_g$ in order for triplet transfer to the acceptor to occur. Additional thermalisation losses will be incurred if triplet excitons possess significantly more energy than needed for transfer.
6. **Stability (resistance to light, chemical and thermal degradation)** – In order to be of practical use in commercial devices, SF chromophores must be sufficiently stable to function over a typical PV device lifespan of 25 years[18].
7. **High Absorption coefficient** – The absorption coefficient of the SF chromophore must be large in order to efficiently harvest high energy photons. This also reduces the necessary thickness of the SF layer, mitigating triplet diffusion losses. The need for a high absorption coefficient can be circumvented through use of an external sensitiser which populates the SF chromophore excited state via Forster Resonance Energy Transfer (FRET). Energy matching between the FRET donor and SF chromophore in this instance will have to be considered to preserve device efficiency.

The total efficiency benefit which can be obtained by the SF process is also affected by the energy difference between the initial singlet exciton and the final triplet products.

In the case that $E_{S1} > 2E_{T1}$, SF is exothermic. Exothermic SF proceeds rapidly since the reaction is both energetically and entropically favourable. SF lifetimes in endothermic systems are on the order of several hundred femtoseconds to several picoseconds, enabling the process to effectively compete with other decay processes. This enables quantitative SF to be achieved in devices. Since the process is exothermic however, the energy difference is lost as waste heat, and triplet excitons resulting from this process are typically low in energy. This can present difficulties in enabling exciton transfer to common acceptors. An example of this is pentacene ($E_{T1}$=0.86 eV), which is poorly matched to one of the most promising acceptors, silicon ($E_g$=1.1 eV).

In the reverse case where $E_{S1} < 2E_{T1}$, SF is endothermic. For endothermic SF, the above benefits and drawbacks are reversed. Energy or charge transfer from triplet excitons produced by SF in endothermic chromophores is possible for a wider range of acceptors since the triplet excitons are higher in energy (i.e. in tetracene $E_{T1}$=1.25 eV). The maximum achievable efficiency gain (45.9%)[16] from utilising an endothermic SF system is also higher than for isoergic systems (41.9%) since this process represents a net energy gain. Despite being energetically unfavourable, endothermic SF may still proceed efficiently due to the inherent entropy gain in generating two triplet excitons from one singlet exciton. However, the process does proceed slower because of this, on the order of tens of picoseconds to several nanoseconds. The slower rate of fission allows competing processes in devices such as singlet exciton transfer, triplet-triplet annihilation (TTA), and diffusion losses to limit the overall SF yield.

Both exothermic and endothermic SF chromophores have been investigated for use in photovoltaic devices. Given the wide variety of possible acceptors currently being explored for SF devices, the merits of both endothermic and exothermic chromophores must be evaluated in each case.

The energetic conditions of SF also require two chromophores to be closely coupled to each other. One method of achieving this is through tight molecular packing of monomer units in polycrystalline films or large single crystals. Chromophores used in this manner undergo intermolecular SF (xSF). SF rates and yields in xSF materials can be highly dependent on the morphology of the film, with domain size and structure playing a significant impact.

Alternatively, a two or more chromophores may be covalently linked in a dimer or polymer structure[30,31]. SF may therefore occur intramolecularly (iSF). Intramolecular systems display significantly higher SF rates than equivalent monomer films since the chromophores are strongly coupled on the same molecule. However, the reverse process of triplet-triplet annihilation is also significantly faster due to this strong coupling and so must be mitigated[32].

### 2.1.2 Requirements for a Singlet Fission Acceptor

For an acceptor material to benefit from SF in a device, it must have the following properties:

1. **Matched energy levels with the chromophore** Analogous to point 5 in Section 2.1.1, the acceptor material must have a conduction band/ LUMO energy aligned with the triplet energy level of the SF chromophore. If this energy level exceeds the triplet energy, triplet exciton transfer will be an endothermic process and may not proceed/proceed slowly with assistance from thermal phonons. If on the other hand the acceptor energy level is significantly smaller than the triplet energy level, transfer will be rapid at the cost of energy lost to thermalisation.

   Any energy losses from energy/charge transfer from triplets derived from SF are doubled compared to an equivalent singlet process as a direct consequence of doubling the number of excitons per incident photon. Appropriate energy matching of the acceptor and SF chromophore is therefore a crucial consideration in developing an efficient SF device.

2. **Efficient triplet exciton transfer mechanisms** The acceptor must be capable of either accepting charges from a triplet exciton or of undergoing resonant energy transfer via the Dexter mechanism.
3. **Efficient charge extraction** The acceptor must facilitate rapid charge transfer away from the interface, both to effectively harvest SF derived photocurrent and to avoid the triplet charge annihilation loss pathway. The presence of trap states in the acceptor due to impurities will inhibit photocurrent extraction and must be avoided.

   Parasitic tail states in disordered semiconductors also adversely impact the $V_{OC}$ of the cell. These states possess reduced energy compared to the HOMO and LUMO band. Carriers quickly thermalise to these tail states, reducing the open circuit voltage by a few hundred meV. This effect is also a loss pathway in non-SF based PV cells. It is important to note, however, that this is a loss process which scales with the number of generated carriers and so will be amplified by the SF process.
4. **Efficient low-band gap power generation** The acceptor material must perform well at low photon energies (i.e. high absorption coefficient, low carrier recombination rates etc) in order to benefit from SF. Since SF acts as a photocurrent multiplier for the device, it will provide the largest performance benefit to acceptor architectures which already possess high efficiencies, but suffer from thermalisation at high photon energies.

### 2.2 Singlet Fission Chromophores

A comprehensive review of SF chromophores was recently published by R. Casilla *et al.* [33], therefore this section will serve as a brief overview of the most relevant chromophores to current and future device studies.

Examples of the categories of chromophores covered by this review are shown in Figure 2 below.

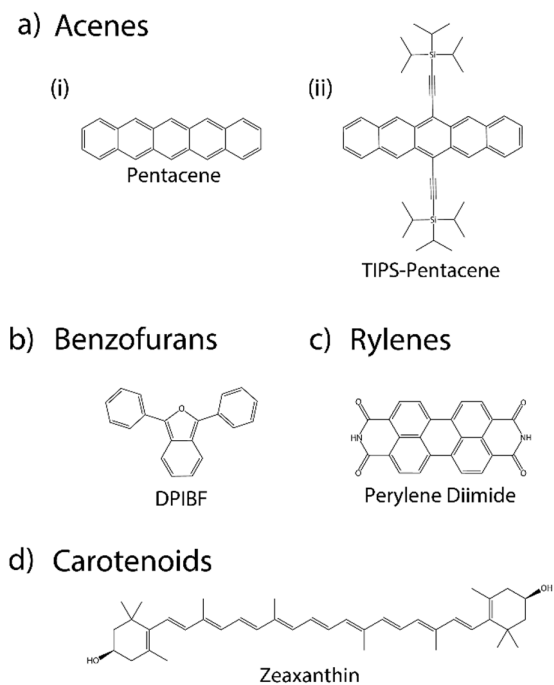

Figure 2: Examples of SF chromophores from the a) acene, b) benzofuran, c) rylene, and d) carotenoid families.

### 2.2.1 Acenes

Acenes are the most extensively studied chromophore type in literature[34,35] beginning with the initial discovery of the SF phenomenon in anthracene crystals in 1965[1]. Their structure consists of fused benzene rings, as shown in Figure 2a for pentacene. Tetracene and pentacene continue to be the focus of most recent literature since they are capable of achieving high yield xSF [19,34,36–40] as well as iSF in dimers [30,41]. In addition, while films of tetracene and pentacene have similar structure, tetracene undergoes slightly endothermic SF and pentacene undergoes exothermic SF enabling direct comparisons of the process.

SF yields in polycrystalline acene films display some resilience to changes in morphology, with phenyl-substituted tetracene films exhibiting a 122% triplet yield despite a loss of long-range order[39]. Disordered pentacene films have been demonstrated to have even greater resilience to molecular disorder, displaying no change in SF rate with the addition of picene or diindenoperylene spacer molecules up to a ratio of 4:1 Spacer:Pentacene[40].

Despite these positive traits there are drawbacks that limit the viability of acenes in commercial devices. The most significant roadblock to implementing acene based SF devices is their poor photostability and oxygen stability. Both pentacene and tetracene readily undergo oxidate under ambient conditions and self-dimerise under photoexcitation [42–45]. The solubility of acenes is also poor, preventing the formation of solution-processed films.

The addition of functionalising groups such as 6,13-Bis(triisopropylsilylethynyl) (TIPS) to acenes as pictured in Figure 2a has been demonstrated to help mitigate these issues[46] by binding to sites in which aggregation could start. One major drawback of this approach however is the effect this has on triplet energy levels. For TIPS functionalised tetracene, the triplet energy level is decreased from 1.25 eV to 1.06 eV[46]. This reduces the maximum band gap which an acceptor may possess in order to harvest triplets from the chromophore. The result of this is a reduction of compatible acceptor materials.

It is clear, therefore, that while acenes will remain useful as model chromophores in the study of fundamental SF processes in solutions, films and devices, they require further modification before they can be considered viable for commercial devices. It is for this reason there has been renewed interest in recent years in discovering alternative SF chromophores.

### 2.2.2 Benzofurans

1,3-Diphenylisobenzofuran (DPIBF), pictured in Figure 2b, is the first successful SF chromophore deliberately engineered for this purpose. It undergoes isoergic SF to produce triplets at 1.42 eV[19,47]. This energy is higher than would be ideal for silicon but would be well suited to GaAs single junction cells (band gap of 1.4 eV).

In films of DPIBF there are two possible structural configuration or polymorphs. The α-DPIBF polymorph undergoes quantitative SF whilst the more thermodynamically stable β-DPIBF configuration forms excimers and so the yield of SF is significantly reduced to 2.0%.

Through the substitution of methyl and butyl groups onto the phenyl rings, bulkier variants of DPIBF were synthesized to examine the role of interchromophore coupling in triplet yields by Dron *et al.*[47]. By increasing the size of the substituted groups, long range order could be disrupted in a similar fashion to the phenyl-substituted tetracene films discussed in section 2.1.1[39]. There is also a clear decrease in the SF rate and yield of the DPIBF variants proportional to the size of the additive[47]. SF in DPIBF therefore is significantly more sensitive to crystal structure than it is for acenes.

The largest issue with DPIBF however is that the stability of the chromophore in oxygen or under illumination is poor even compared with non-functionalised acenes[19]. Thus, DPIBF is also not suitable for use in practical SF devices but is a successful demonstration of engineered SF chromophores.

### 2.2.3 Rylenes

Rylenes have attracted some interest in recent literature due to their increased stability in comparison with acenes and large absorption coefficients[48,49]. Perylenes are already widely used as industrial pigments due to their high thermal, chemical and photostability as well as their strong absorption and fluorescence in the visible wavelength range[50,51].

Perylene diimide (PDI), pictured in Figure 2c, has been shown to undergo endothermic SF with an energy surplus of the $T_1+T_1$ state larger than the $S_1$ state $2E_{T1} - E_{S1} = 0.2$-0.3 eV[48,49]. This high level of endothermicity means that the rate of SF is slow in comparison to exoergic systems with a measured rate constant of 180ps – 3.8ns in films [48,49]. Despite this, SF can proceed efficiently with a 140% exciton yield likely due to the entropy increase inherent in exciton multiplication[48].

In 2019 Conrad-Burton *et al.* attempted to shift the energy levels of PDI through molecular contortion[49]. By adding side chains to PDI to produce a derivative named PDI-B they applied a longitudinal contortion to the molecule. This contortion was calculated by DFT to increase the two-electron exchange energy and therefore lower the singlet energy by 0.1 eV and the triplet energy by 0.2 eV[49]. Consequently, this would increase the singlet-triplet energy gap sufficiently for SF to become isoenergetic.

This DFT result was indirectly confirmed by transient absorption spectroscopy on films of PDI-B which showed an SF lifetime reduction to 2.5 ps. This improvement of nearly two orders of magnitude was also accompanied by a 2-3 order of magnitude reduction in triplet lifetime to 160 ps in PDI-B[49].

Terrylene diimide (TDI) has similarly been shown to undergo SF. Unlike PDI, TDI has a $T_1$ energy of 0.77 eV and $S_1$ energy resulting in exoergic SF[52]. The $T_1$ yield is quantitative in TDI thin films and proceeds via a fast (~3.0 ps) and slow (~30 ps) rate constant. The triplets produced by this process have a lifetime of approximately 50 ns [52].

Attempts to solubilise TDI using tert-butyl substitutions have resulted in a variant that maintains quantitative SF whilst being weakly endothermic by 0.13 eV[53]. The $T_1$ energy of this variant is 1.1 eV which is well matched to the band gap of silicon. The SF timescale for this variant has increased to 120-320 ps but remains competitive with other $S_1$ decay pathways, such as fluorescence[53].

Therefore, as a result of their high SF yields, stability and useful range of $T_1$ energies rylenes are promising candidates for realising practical SF devices.

### 2.2.4 Carotenoids

Carotenoids are a group of organic molecules found naturally in biological systems. They play a role in both light harvesting and photoprotection in photosynthetic organisms such as plants and bacteria[7,8,54–56].

In 1980, SF was observed in the carotenoid spirilloxanthin, present in the photosynthetic antenna complexes present in bacteria[8]. Magnetic field dependent fluorescence measurements confirmed the presence of high triplet quantum yields (~30%), which were quenched in the presence of a large magnetic field (~0.6T). Carotenoids are typically found paired with chlorophyll in the chloroplasts of photosynthetic organisms. The triplets produced by SF in carotenoids are able to inhibit the harmful photooxidation of molecular oxygen to singlet oxygen[7,54–57]. Whilst carotenoids do also play a role in sensitising chloroplasts to light in the 450-570nm range, this has been shown to be due to singlet-singlet energy transfer.

Since this initial observation, SF has been discovered in numerous other carotenoids, including zeaxanthin (Figure 2d), β-carotene and astaxanthin. SF in carotenoids is dependent on morphology, with aggregation sites in zeaxanthin demonstrated to show significantly higher triplet yields (90-200%)[58,59] compared to the isolated monomer (0.2%). This high sensitivity to molecular contortion may enable organisms to regulate the SF reactivity of carotenes using binding proteins in response to oxidative stress[59].

In the field of photovoltaics, carotenes are of interest due to their fast SF rates and unique reaction energetics. Carotenes display some of the fastest intermolecular SF reaction rates recorded in literature to date, with a SF time constant of <70fs measured in astaxanthin aggregates[59]. Fission appears to proceed from the $S_1$ state directly to the $(T_1+T_1)$ state without an intermediate. [57,59] These kinetics are typically associated with intramolecular SF in acenes, and so further studies of these chromophores may provide some insight into the underlying photophysics of the SF process[59].

# 3. SINGLET FISSION PHOTOVOLTAIC DEVICES
## 3.1 Is Singlet Fission Playing a Role?

Effectively assessing the progress of SF-based photovoltaics is less straightforward than judging the progress of commercially successful photovoltaic devices. Two steady-state techniques for quantifying the contribution of SF to device performance will be discussed at the beginning of this section before moving on to a discussion of SF devices.

Time-resolved optical pump-probe spectroscopy can be very useful in assessing the effect of SF on photocurrent, but is not suitable for all device architectures and is not generally representative of device performance under solar illumination conditions due to high pump fluences. It may provide information about singlet and triplet kinetics[30] and triplet charge transfer across an interface[60–63] but is not generally used as a standalone device characterisation method. As such, we focus on steady-state methods in Sections 3.1.1 and 3.1.2 below.

### 3.1.1 Quantum Efficiency Methods

A common method of determining the presence of SF in a device is through calculating its external and internal quantum efficiency. The external quantum efficiency (EQE) of a cell is the ratio (given as a percentage) of the number of electrons produced by the device compared to the number of incident photons used to illuminate the device. In devices that do not utilise the SF process (or an alternate form of multiple exciton generation), the maximum possible EQE is 100%. Since the SF process produces two photoelectrons per photon, the maximum EQE will instead be 200% in a SF device. An EQE>100% is therefore an unambiguous indication that SF is occurring in a photovoltaic device.

A typical measurement apparatus for determining the EQE of a photovoltaic device utilises a white light source of known intensity and a monochromator for wavelength selection. A beam splitter is used to separate some of this light for photon flux measurements with a reference photodiode, whilst the remainder of the light is used to illuminate the device. A chopper is placed before the beam splitter in the light path so that a lock-in amplifier can be used to measure the photocurrent from both the reference photodiode and the photovoltaic device.

Since the photon energy dependent EQE of the reference photodiode is known, the EQE of the photovoltaic cell can be calculated by:

$$EQE_{cell}(E_\gamma) = \frac{I_{cell}}{I_{ref}} EQE_{ref}(E_\gamma)$$

For an ideal SF cell, the EQE will be 200% for $E_\gamma > E_{S1}$. Note that this will only occur within the absorption range of the SF chromophore, and photons below this energy may only achieve a maximum EQE of 100%. In practice however, factors such as cell reflectivity, absorption by the encapsulant, and the relative absorbance of the SF chromophore and acceptor layer will affect the final cell EQE, potentially reducing it below 100% even in the case that SF is occurring efficiently.

In order to determine if this is the case, the EQE of the cell must be normalised with respect to the absorbance of the cell to find the internal quantum efficiency (IQE). This IQE measurement therefore provides a measurement of the number of photoelectrons generated per absorbed photon. Through separately calculating the absorbance of each layer, the IQE of each active layer can be modelled from the device IQE, providing insight into whether SF is occurring efficiently.

Given accurate absorbance measurements, an IQE>100% also represents unambiguous verification of SF. In nanostructured devices however, scattering reduces the accuracy of absorbance measurements. This can limit the utility of IQE measurements in determining the SF efficiency in these devices. IQE measurements are still qualitatively useful in this case since the presence of chromophore absorption features in the IQE confirms that exciton transfer is taking place.

### 3.1.2 Magnetic field dependent photoluminescence and photocurrent measurements

Although an IQE>100% is a definitive demonstration of SF contribution to cell operation, the development of new architectures or materials often requires the impact of SF to be characterised far from these optimal conditions, and where IQE can be much lower than 100%. The IQE spectrum may show the absorption signature of SF materials; however, in slower, endothermic SF systems this may be due to FRET from $S_1$ before fission occurs. It is important that techniques which can unambiguously demonstrate SF in these regimes are available.

One approach is to use the effect of magnetic fields on these systems. The efficacy of the conjugate processes to SF, TTA via the singlet channel ($TTA_S$) to produce either $S_1+S_0$ or $S_0+S_0$, or TTA via the triplet channel ($TTA_T$) to produce $T^* + S_0$, is magnetic-field-dependent. The former may be monitored via the emission from the $S_1$ state. This can be exploited to ascertain whether a solar cell is generating current via the SF channel.

The effect arises from the fact that magnetic fields alter the distribution of singlet character amongst the 9 triplet pair states, and was first explained by Merrifield[2,6,64,65] and then revisited by Bardeen and coworkers[66–69]. See these references for details on the specifics of kinetic and spin models to calculate quantitative effects. In this section we reframe these previous formalisms to provide an intuitive way in which to understand the qualitative observations and to explain the complexity of the phenomenon.

Parallel Chromophores

In this section we consider the simplified case of two parallel chromophores and the Hamiltonian described in Ref [6]. We will first consider the zero-field and high-field (i.e. where the Zeeman term dominates over the zero-field term in the spin Hamiltonian) cases.

At zero field, we represent individual triplets in the basis $|x\rangle$, $|y\rangle$, and $|z\rangle$, corresponding to triplets which precess about the principal axes of the zero-field splitting tensor, $\mathbf{D}$, (usually aligned with the molecular principal axes). The singlet state (i.e. the eigenvector of the $\widehat{S^2}$ operator with eigenvalue $S(S + 1) = 0$) is a linear combination of three product states,

$$^1TT = \frac{1}{\sqrt{3}}(|xx\rangle + |yy\rangle + |zz\rangle)$$

At high field, triplet states $|-\rangle$, $|+\rangle$, and $|0\rangle$ are described relative to the applied magnetic field, and are respectively anti-parallel, parallel, and precess about the field vector. In this regime, only two spin eigenstates $|00\rangle$ and $(|+-\rangle + |-+\rangle)$ comprise the singlet eigenvector,

$$^1TT = \frac{1}{\sqrt{3}}(|00\rangle - |+-\rangle - |-+\rangle)$$

At intermediate fields the number of triplet pair states over which $^1TT$ character is distributed varies (Figure 3). This distribution of singlet character affects the efficacy of TTA$_S$ and TTA$_T$. We consider both spin-correlated and spin-uncorrelated triplet pairs in the phenomenological explanation below.

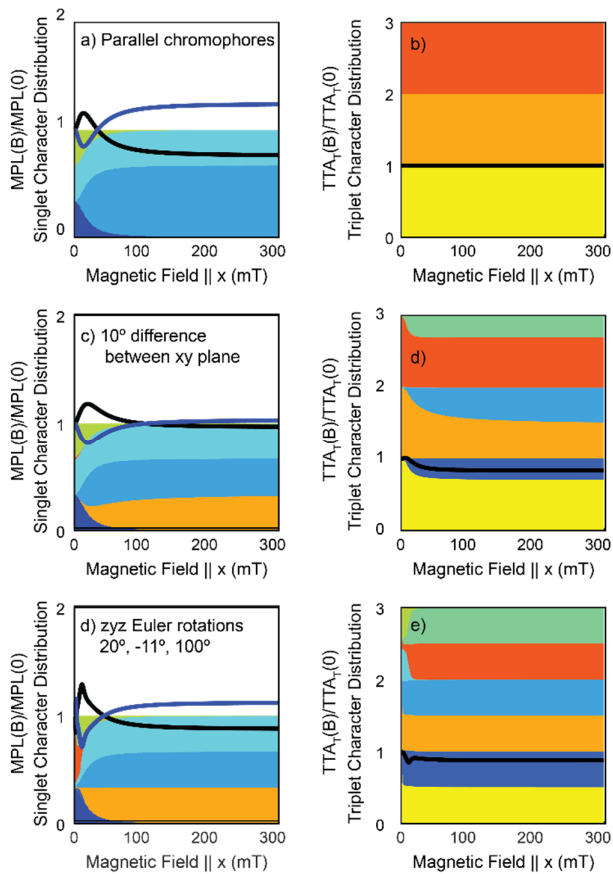

Figure 3: Distribution of singlet and triplet character over product states as a function of magnetic field strength in the molecular x-direction for (a,b) two parallel chromophores, (c,d) chromophores with a 10° angle between the xy plane of the chromophores, and (e,f) using Euler rotations with angles 20°, -11°, -100°. Each solid block of color represents a different triplet pair state. Schematic of the normalized dependence of PL from TTA$_S$ of uncorrelated (black line) and correlated (blue line) triplet pairs are also shown in (a,c,e) using the formalism in Ref 6 and $k_1 = 1.7 \times 10^9$ s$^{-1}$, $k_T = 1.7 \times 10^6$ s$^{-1}$, $k_S = 1.1 \times 10^{10}$ s$^{-1}$, $k_{-1} = 2.8 \times 10^9$ s$^{-1}$. The efficacy of TTA$_T$ of uncorrelated triplets is shown in black in (b,d,f). Parameters used in this simulation are $D = 1000$ MHz, $E = 50$ MHz, $g = g_e = 2.0023$. A small inter-triplet exchange energy of 1 MHz was incorporated to break degeneracies.

There are three potential fates of colliding triplets (T$_1$+T$_1$): TTA$_S$, TTA$_T$, and scattering to remain in T$_1$+T$_1$. TTA to produce Q$_1$+S$_0$ (TTA$_Q$) is not observed since a quintet state residing on a single chromophore is energetically inaccessible. We note that, in the case of parallel chromophores, the triplet pair spin wavefunctions of the singlet and quintets ($^1TT$ and $^5TT$) are symmetric with respect to exchange of a triplet, whereas the triplet spin wavefunctions ($^3$TT) are antisymmetric with respect to triplet exchange,[6] which will become important below.

Considering the zero-field case, SF produces a triplet pair in the exchange-coupled $^1$(TT) state. These triplets may then dissociate into uncoupled triplets ($T_1+T_1$), however their spins remain correlated for many microseconds[70]. The important thing to consider in this scenario is that the ($T_1+T_1$) state begins with 100% singlet character that can only be *diminished* through decay to other states.

At zero-field, the singlet, $^1TT = 1/\sqrt{3}\,(|xx\rangle + |yy\rangle + |zz\rangle)$, evolves and, since the $|xx\rangle$, $|yy\rangle$, and $|zz\rangle$ product states have different energies their relative phases vary. This yields character of other spin states, and has been observed as quantum beating in the delayed fluorescence[71,72] with three characteristic frequencies corresponding to the three energy differences: $E_{xx} - E_{yy}$, $E_{yy} - E_{zz}$, and $E_{xx} - E_{zz}$. That is there are three spin pathways out of the singlet state.

By contrast, at high-field, there are only two spin eigenstates in $^1TT = 1/\sqrt{3}\,(|00\rangle + (|+-\rangle + |-+\rangle))$. As $|+-\rangle$ and $|-+\rangle$ are degenerate, there is only one energy difference of interest, $E_{00} - E_{+-} = E_{00} - E_{-+}$, and this has also been observed as a single quantum beat frequency[73]. That is, there is only one spin pathway out of the singlet state.

The net effect of this on the MPL from correlated triplets is that there are more pathways to diminishing singlet character of correlated triplet pairs at zero-field, as opposed to high field, giving rise to the trend shown in the blue line in Figure 3a.

The reverse effect is observed for uncorrelated triplet pairs which have an equal probability of occupying each of the nine pair states. Any triplet-triplet collision of triplet pairs with symmetric spin wavefunction will result in either TTA$_S$, scattering back to $T_1+T_1$ or decay to $T_1+S_0$. Importantly, $T_1+T_1$ may subsequently recollide and result in TTA$_S$. Since uncorrelated triplet pairs are equally likely to occupy each of the nine product states, TTA$_S$ is enhanced by the number of pair states with singlet character and MPL is diminished at high-field (black line in Figure 3a).

The net effect of MPL on correlated and uncorrelated triplet pairs is best observed in Figure 4, from Ref.[66]. Here, at times $\lesssim 20$ ns, the high-field case reduces the number of pair states with singlet character, enhancing PL from spin correlated triplets. However, at times $\gtrsim 20$ns uncorrelated triplets are colliding and an applied field reduces MPL. Similar observations have been made in other SF systems[74].

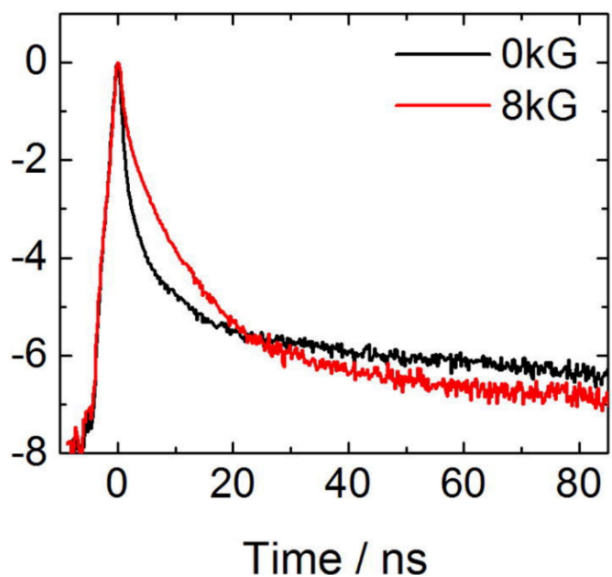

Figure 4: Fluorescence decay in polycrystalline tetracene films under zero field (black) and 8 kG external field (red). Reproduced from Burdett *et al.*[66] Copyright (2021) with permission from Elsevier.

<u>The General Case</u>

The phenomenological description in this section is useful for understanding how the distribution of singlet character across pair states varies under a magnetic field and that this can enhance or diminish MPL depending on the spin correlation of interacting triplets. However, *qualitative* differences will be observed for systems with differing relative alignment of chromophores undergoing TTA$_S$ and fluctuations in exchange coupling[75].

Importantly, under the special conditions above, the efficacy of TTA$_T$ is independent of magnetic field[6,64] (Figure 3b). This breaks down for non-parallel chromophore pairs, where the SF-generated $^1$TT state may evolve to obtain both $^3$TT and $^5$TT character[75]. Similarly, uncorrelated triplets which collide to form $^1$TT character may evolve to the $^3$TT state and undergo TTA$_T$[76]. Even a 10º difference in the orientation of chromophores can give rise to very different results (Figure 3(c,d)). Lower symmetries yield more complex results (Figure 3(e,f)).

An inhomogeneous sample will therefore have many possible magnetic-field dependencies. As such, the useful measure for a device physicist when assessing the efficacy of triplet augmentation of solar cell current is to compare the time-integrated MPL effect to a magnetic-field-dependent photocurrent (MPC) measurement. If the parity of the MPC is opposed to the MPL one can conclude that the photocurrent is positively correlated with triplet population. (An important departure from this rule exists for certain device architectures wherein triplet transfer may passivate the acceptor material, improving its performance, but not augmenting current, and further measurements may be needed)[77].

Finally, the above argument is useful when TTA$_S$ to $S_0 + S_1$ is an active channel, and therefore MPL and MPC may be compared. This is generally not the case for exothermic SF systems wherein this process is energetically inaccessible. However, if the TTA$_T$ channel is open, TTA$_S$ to $S_0 + S_0$, or triplet-radical interactions occur, an MPC effect will still be observed, and one can conclude that triplets are playing a role in the device. Combining these properties, we construct the flowchart below to assist with identifying when triplets derived from SF are enhancing PV device operation.

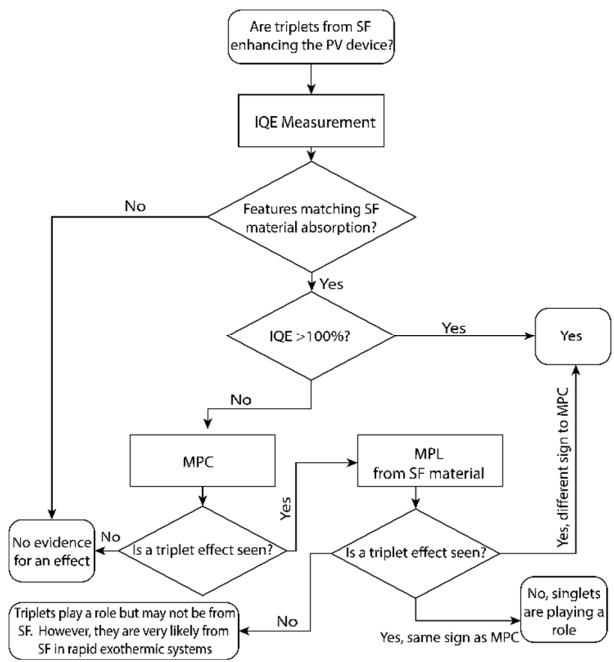

Figure 5: Flowchart describing the process for determining if SF is improving photocurrent in a device.

## 3.2 Categorization of Singlet Fission Devices

There have been many attempts to produce SF solar cells using different materials and architectures. All devices rely of charges from triplet excitons – either via Dexter energy transfer or charge transfer – at an interface. As such, we categorise these approaches by the interface material being either organic, inorganic quantum dots, mesoscopic semiconductors, perovskites, or crystalline silicon.

### 3.2.1 Organic Photovoltaic Devices

The active layer of an organic photovoltaic device typically consists of a heterojunction containing an electron donor and electron acceptor material. At this interface, photoexcited excitons undergo electron transfer to form an interfacial charge transfer state, followed by dissociation of the bound charge transfer state into an electron in the LUMO of the acceptor and a hole in the HOMO of the donor.[38,78–82]

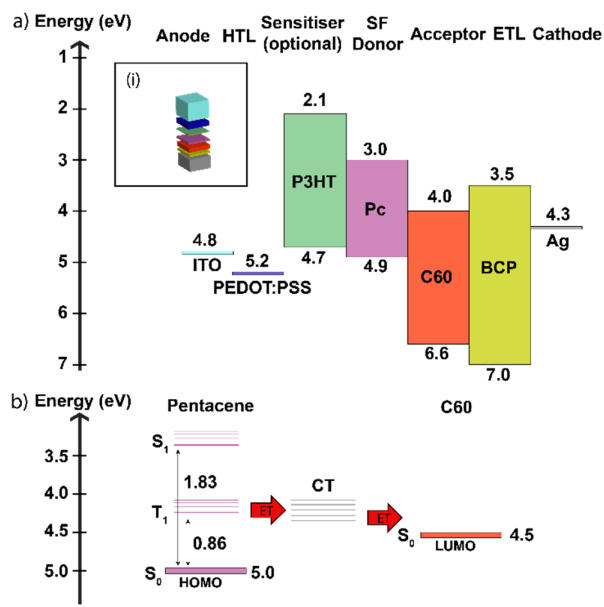

Figure 6: a) Exploded view of planar OPV structure. b) Energy level diagram of the Pc/C60 planar OPV structure used by Congreve et al [38]. c) State diagram of the Donor/Acceptor interface in a Pc/ $C_{60}$ OPV. Energy levels of layers b) and states c) are given in eV.

Since SF occurs only in organic chromophores, this was a logical starting point for device implementation. A singlet state in the SF material is populated either by direct photoexcitation[83] or through Forster Resonance Energy Transfer (FRET) from an external sensitiser [38,80]. Triplet excitons are formed within the SF layer and then must diffuse towards the interface where they can then form a charge transfer state.

The simplest SF-OPV device architecture is the planar heterojunction device shown in Figure 6a. This consists of a glass substrate with a transparent hole extracting front electrode such as ITO contacting a donor layer. An acceptor layer is deposited atop the donor layer, forming a neat heterojunction. Electrons are extracted from the device using a rear electrode such as Al (due to favourable energy level alignment), the reflectivity of which also serves to increase the optical path length in the device and therefore the absorption. Exciton blocking and hole/electron transport layers are often inserted between their respective electrodes and the active layers for improved charge collection efficiency[38,80,83–85].

The first such device consisted of a pentacene/$C_{60}$ donor-acceptor interface with BCP inserted between the Al and $C_{60}$ layer as an exciton blocking layer. In two papers, Yoo *et al.* identified that EQE and IQE of this cell was high, particularly where pentacene was absorbing, but did not consider SF in their analysis, instead assuming that the efficiency of charge separation and collection in the cell was close to unity[83,85].

It is likely, however, given the behaviour of pentacene/$C_{60}$ junctions observed in later work,[84] that this high quantum efficiency was in part due to the harvesting of triplets generated through SF in the pentacene film. Through a series of studies, the Baldo Group at MIT demonstrated high IQE and EQE due to SF in pentacene/$C_{60}$ devices. In 2009, Lee et al. sought to exploit SF in Pc/$C_{60}$ heterojunctions by fabricating a variation of the planar heterojunction device consisting of 30 bilayers, each made up of 2nm of pentacene and 1 nm of $C_{60}$.[84] This structure aimed to reduce exciton diffusion losses by reducing the distance needed to travel by excitons from the pentacene to the heterojunction interface. Multiple bilayers were used to compensate for the low absorption of the thin pentacene layers. This device had to be operated at a high reverse bias (3.5V) to overcome charge trapping at the layer interfaces and so was not suitable as a functional photovoltaic cell architecture for power generation. However, this does yield an effective photodetector architecture[84,86] The internal quantum efficiency for pentacene in this device was calculated to be (128 ± 2)%, indicating that SF was enhancing the efficiency of the photodetector. Magnetic field dependent photocurrent measurements confirmed that pentacene was contributing to an exciton yield of (145 ± 7)% in the device[84].

A key result of this body of work was that the triplet exciton diffusion length and the effect of chromophore layer thickness on the SF yield for pentacene[38] and tetracene[81] devices. For both SF chromophores, the multilayer architecture was used to increase charge collection efficiency in layers <5nm during magnetic field dependent current and fluorescence measurements. Pentacene was demonstrated to undergo complete SF in layers thicker than 15 nm, whilst tetracene required layers more than 100nm in thickness. Pentacene- and tetracene-based cells exhibited a decrease in IQE due to diffusion losses as the layer thickness increased. For pentacene, however, the earlier peak in SF efficiency meant that a higher IQE of (160+/-10)%[38] was achieved (at 15 nm) compared to tetracene, which achieved a peak efficiency of (127+/-18)%[81] at 25 nm layer thicknesses. Since this peak occurs well before quantitative SF yields are achieved, it is clear that the slow rate of SF in tetracene significantly limits the potential efficiency gain possible in this device.

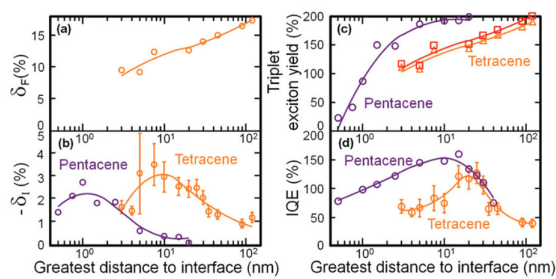

Figure 7: a) The tetracene fluorescence change under a magnetic field, b) Photocurrent change under a magnetic field for both tetracene and pentacene devices, c) Calculated triplet exciton yield for tetracene and pentacene devices, for tetracene MPL only (red squares) and MPL & MPC derived yields (orange triangles) are shown, d) IQE for tetracene and pentacene devices. All factors are plotted as a function of interface distance (SF layer thickness). Reprinted from Wu *et al.* [81], Copyright (2021), with the permission of AIP Publishing.

SF-OPV bulk-heterojunction OPV has also been demonstrated by the Baldo Group, as shown in Figure 8. The BHJ utilises a blended donor-acceptor layer instead of distinct donor and acceptor layers. This blended junction maximises the surface area of the donor-acceptor interface, improving charge separation efficiency compared to a planar device architecture. This architecture also has the advantage of being solution processable, enabling mass production of devices using roll-to roll printing techniques.

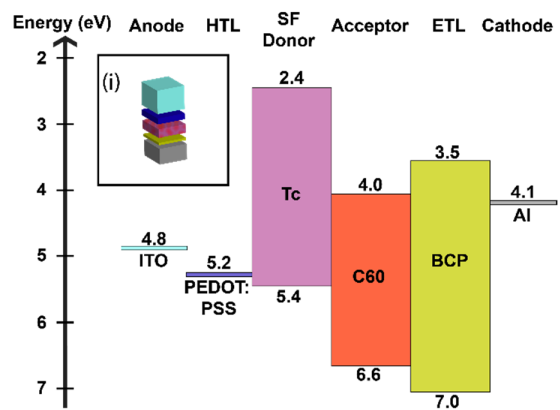

Figure 8: Bulk heterojunction architecture (left) with energy level diagram (right) for the device structure by P. Jadhav et al.[86].

It is for these reasons that the BHJ architecture is dominant in commercial OPVs. The first attempt to integrate the SF process into the BHJ architecture was by Jadhav et al. in 2011[86].The device architecture and its corresponding energy level diagram is shown in Figure 8, with tetracene and $C_{60}$ comprising the active layer.

Magnetic-field-dependent photocurrent measurements were conducted for varying ratios of Tc:$C_{60}$, demonstrating relative stability in the effect down to 20% Tc concentration. From this result Jadhav et al. surmised that the tetracene SF rate in the BHJ active layer was significantly faster than the exciton dissociation rate [86]. Whilst this result differs from what would be expected in a uniform tetracene film[81], the presence of aggregates in the junction with enhanced SF coupling provides a likely explanation. Later investigations of polycrystalline tetracene films support this assertion, with smaller crystal grain sizes leading to increased SF rates over large grains in the stable Tc I polymorph[87].

A later investigation by Thompson et al. using pentacene based BHJ cells focused on determining the impact of the triplet charge annihilation (TCA) loss mechanism[88]. In this interaction, the SF-derived triplet exciton annihilates to the ground state upon interaction with a polaron. Whilst this loss mechanism is likely to be present in any SF device, the presence of a high surface area heterojunction throughout the bulk of the active layer in a BHJ provides significantly more opportunities for triplet-charge interactions than a planar junction. Different PC:C60 blends in OPVs in addition to planar and multilayer heterojunction photodiodes were discussed in this study using magnetic field dependent photocurrent spectroscopy. TCA was demonstrated to be a significant loss factor in the BHJ architecture, accounting for a loss of approximately 50% of all triplet excitons generated in the pentacene for a BHJ cell with a 2:1 ratio of Pc:$C_{60}$. With the application of a 2V reverse bias to the cell, a decrease of TCA from 50% to under 25% was observed due to more efficient charge extraction, highlighting the need for rapid charge collection in SF devices. BHJ architectures were found to be most susceptible to TCA.

Whilst the creation of high efficiency SF-based OPVs has not been a significant focus of the field as of 2020, there are potential benefits to such a device. Preliminary investigations into the implementation of tetracene into the bulk heterojunction architecture suggest that tetracene forms aggregates with more favourable SF geometry than neat films. This may assist in overcoming the IQE limits observed in planar heterojunction devices caused by triplet diffusion losses since the donor acceptor distance in small in blended films. Increased triplet charge annihilation however is a factor which will need to be overcome to enable SF to provide a net benefit to device performance. Two other factors which limit the potential of SF OPVs are: the energy cost of exciton dissociation and the presence of parasitic 'tail states' due to the disordered nature of the organic semiconductor. Both of these factors exist in all organic photovoltaic devices but their consequences are compounded by the SF process. In the case of dissociation losses, the need to dissociate two triplet excitons rather than one singlet exciton doubles the size of this loss pathway. The formation of the pentacene/C60 triplet charge transfer state for example has been demonstrated by Willems *et al.*[89] to be endothermic by ~0.1eV and therefore will act to reduce the open circuit voltage $V_{OC}$ of a pentacene/C60 device. Efforts to limit the total $V_{OC}$ losses in OPVs will be necessary if SF-OPVs are to be a viable device for power generation. At the time of this review however, OPVs remain a useful platform for the examination of SF and exciton transport behaviour in devices.

### 3.2.2 Inorganic/Organic Quantum Dot Hybrid Devices

Hybrid devices formed from pairing a SF chromophore with low band gap quantum dots (or nanocrystals) were trialled as early potential candidate architectures for the utilisation of SF. Quantum dots are semiconductor nanoparticles with radii smaller than the exciton Bohr radius (typically a few nm), wherein quantum confinement effects result in discrete energy levels similar to those seen in atoms or molecules. Since electrons are confined to this small radius, the tuneable energy levels of quantum dots enable precise energy level matching with a SF chromophore. Quantum dots and solubilised SF materials are also both solution processable[90], and so have the potential for significantly reduced energy and financial cost per molecule compared with conventional silicon solar cells if produced at commercial scales. This may offer a significant further reduction in the cost and emissions intensity per watt generated of solar power. Like OPVs, quantum dot solar cells can also be fabricated on lightweight, flexible substrates enabling a wider range of applications.

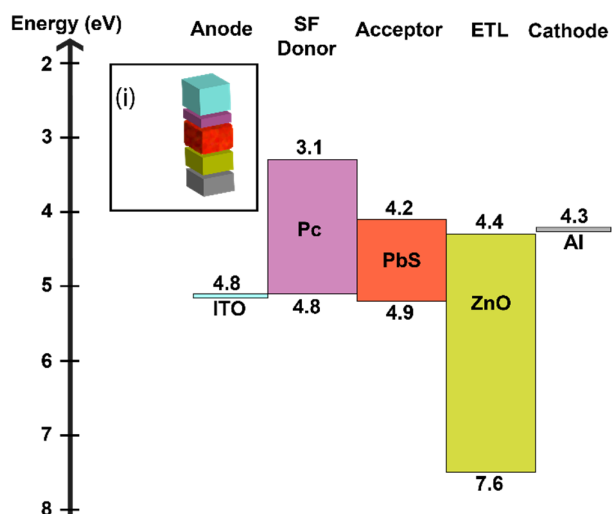

Figure 9: SF-sensitised quantum dot-based cell architecture(left), with energy level diagram (right) by Ehrler *et al.*[92].

The hybrid devices which have been produced in literature share a similar architecture to the planar heterojunction OPV devices in the previous section. A transparent top electrode (ITO) is used as the hole extraction layer. In current literature, pentacene and TIPS-pentacene have been used as the SF chromophore[90–92]. The inorganic quantum dots act both as electron acceptors for triplets generated in the pentacene and as low-band gap absorbers. Since pentacene is transparent to infra-red light, the bilayer design allows high energy photons to be absorbed in the pentacene, whilst allowing the quantum dots to effectively absorber low energy photons.

Both PbS and PbSe nanocrystals have been trialled as nanocrystal materials, with surface ligands such as BDT, TBAI and EDT added to crosslink the nanocrystals in order to form and interface aligned to the pentacene. The addition of an electron transport layer such as ZnO or TiO$_2$ has been shown to significantly improve the efficiency of QD hybrid devices, i.e. the first SF sensitised QD study showed an improvement from 0.16% PCE to 0.85% upon addition of a 100nm ZnO nanocrystal layer[92]. A 1 nm thick LiF layer between the nanocrystal layer and electrode was also demonstrated to maintain device performance under constant illumination, due to reduced interfacial charge trapping.

The tuneable nature of QD acceptors has enabled devices to be used as a method of estimating the triplet energy of SF chromophores. Ehrler *et al.* performed QE measurements of pentacene based cells with PbSe nanocrystals possessing a range of band gap energies from 0.67-1.20 eV[91]. The EQE spectra showed a drop in pentacene contribution between 1.08 eV and 1.20 eV, allowing an upper and lower bound to be established for the pentacene triplet energy. After correcting for the photocurrent onset voltage, the range for pentacene triplet energies in devices was found to be 0.85 eV<$T_1$<1.0eV, consistent with previously calculated values in films.

QD hybrid devices are among the most successful efforts to produce an efficient SF based photovoltaic device to date. In a device with TIPS-Pc as the SF chromophore and a 1.25eV PbS QD acceptor, a PCE of 4.8% was recorded with a $J_{sc}$ of 19.8 mAcm$^{-2}$ and $V_{OC}$ of 0.59 V [90]. QE measurements confirmed that the IQE of TIPS-Pc was (160±40)% and therefore evidenced that efficient harvesting of triplet excitons occurs in this device structure. However, in a similar vein to SF OPV devices, QD hybrids have several hurdles to overcome.

### 3.2.3 Dye-sensitised solar cells

Dye-sensitised solar cells (DSSCs) utilise a high band gap metallic oxide semiconductor as the charge extraction layer for a sensitiser molecule with a large absorption coefficient. Since charge extraction occurs at the dye/semiconductor interface, it is common to utilise a mesoporous semiconductor layer composed of sintered nanoparticles. The semiconductor layer is soaked in the dye and dried to form a monolayer of dye molecules adsorbed to the porous semiconductor surface, maximising light absorption and charge collection area. A liquid redox electrolyte is often used as an electron donor to replenish the dye after electron injection.

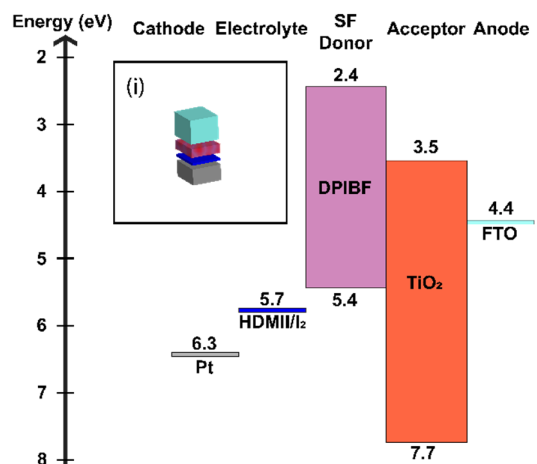

Figure 10: Singlet fission DSSC Structure (i) and energy level diagram for the device structure by Schrauben *et al*[93].

In the case of SF devices, DSSCs provide a useful platform to investigate charge extraction at the organic/inorganic interface as a result both the large charge collection area and efficient charge transport compared to OPVs. This device structure has the advantage of both minimising exciton diffusion losses due to the short distance to a charge transfer interface and reducing triplet charge annihilation due to more rapid charge transport in comparison to OPVs.

The first device SF-based DSSC was produced in 2015 by Schrauben *et al.*[93], utilising the small molecule SF chromophore DPIBF as the donor and $TiO_2$ as the acceptor. This chromophore displays high room temperature SF yields at room temperature (~140%) in its α polymorph and significantly lower yields in its more stable β polymorph (~10%). Due to the difficulties in controlling morphology in DSSCs, EQE and IQE values did not exceed 100% (60-70% IQE calculated) and so were insufficient alone to determine the triplet transfer efficiency.

Kinetic modelling suggested a majority singlet contribution to the photocurrent due to a rapid charge injection rate (<1 ps) which significantly outpaced the SF rate of α-DPIBF (30 ps). Upon addition of a $ZrO_2$ barrier layer, electron injection was able to be sufficiently slowed for triplet formation via SF to occur. This was observed by examining the how the photocurrent varied as a function of barrier layer thickness, where a discontinuity was seen in the overall downwards trend when the triplet charge contribution became a significant contributor to device photocurrent. A comparison of this trend with the expected result from kinetic modelling is shown in Figure 11[93]. The overall PCE of this device was 1.1% with $V_{OC}$=0.51V and $J_{SC}$=3.3 mAcm$^{-2}$.

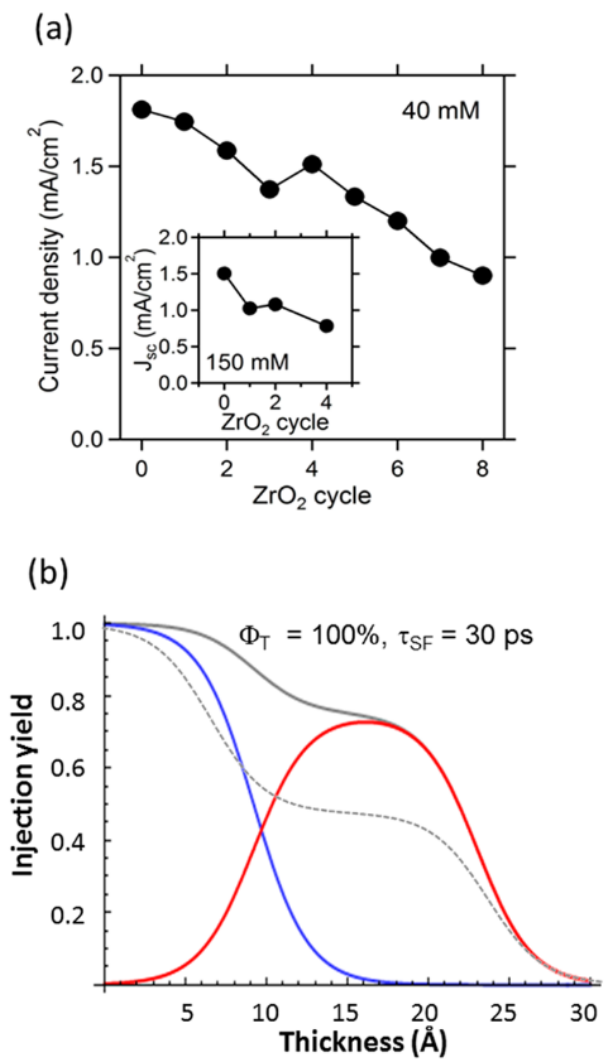

Figure 11: a) JSC vs number of ZrO$_2$ deposition cycles for DPIBF sensitised cells, b) Predicted injection yield from kinetic simulations. Total exciton injection yields are shown, with the blue representing S$_1$ injection, red representing T$_1$ injection and grey representing total injection (solid line assuming 100% T$_1$ and dashed assuming 50% T$_1$ injection efficiencies). Figure from Schrauben, J. N. *et al.*[93] , Copyright (2021), Reprinted (adapted) with permission from American Chemical Society.

A version with a modified version of DPIBF with a saturated hydrocarbon chain to act as a spacer, terminated in a carboxylic acid group to allow covalent bonding to the TiO$_2$ has also been fabricated. Whilst the overall efficiency was lower than with pure DPIBF (0.45% PCE), the device IQE measurements were consistent with kinetic simulations and indicated 150% T$_1$ transfer efficiency. An increase in photocurrent was noted with a thinner ZrO$_2$ layer than the unmodified version (1-2 deposition cycles vs 3-4) and accompanied a measured increase in IQE over a device without the barrier layer. The drop in performance compared to unmodified DPIBF was a result of a lower photocurrent (1.4 mAcm$^{-2}$), attributed to poor dye loading.

A later 2018 investigation of DPIBF in DSSCs by Banerjee *et al.*[94] replicated this behaviour in both TiO$_2$ and ZrO$_2$ based DSSCs with alumina (Al$_2$O$_3$) barrier layers. Emission from the DPIBF decreased as a function of increased dye loading during time-resolved emission spectroscopy. This result was attributed to reduced singlet formation due to increased intermolecular interactions to form either a triplet state via SF or an excimer state. Long-lived emission consistent with the presence of an excimer state was observed. Features consistent with long-lived triplets were present in ns-TA measurements although the yield could not be calculated accurately using fs-TA since the spectra were ambiguous and likely contained multiple species such as CT states.

Magnetic field dependent fluorescence and photocurrent measurements up to 0.55 T did not show a response which indicated the presence of SF. As discussed above, this result does not conclusively prove that SF is not occurring in this system since the magnetic field response of DPIBF is currently unknown[19,94]. An investigation into the magnetic field response (including determination of the ZFS parameters) would be needed for this to be verified. Banerjee *et al.* [94] also showed that a bilayer of DPIBF and a ruthenium dye could be formed via a $Zn^{2+}$ linking ion and carboxylic acid functional groups on each molecule. Whilst the nature of the interface is not yet known, emission from the Ru dye was completely quenched in this interface, indicating 100% efficient triplet energy transfer to the DPIBF. This process may allow both a SF capable and low-band gap dye to be bonded to the DSSC without sacrificing dye loading of either compound.

Efforts to incorporate acenes into DSSCs were also undertaken by Kunzmann *et al.* in 2018[62,63], using pentacene derivatives on $TiO_2$, ZnO and InZnO substrates. Doping the metallic oxide substrate was demonstrated to be an effective means of permitting greater triplet exciton transfer through lowering the quasi-Fermi level of the semiconductor. For investigations of pentacene-based monomers, a record device efficiency of 1.52% was obtained through optimisation of the Li+ ion concentration. Due to the negative impacts on electrolyte efficiency and the $V_{OC}$ of the device, there are limitations to the degree of effective tuning possible. The presence of triplet excitons was confirmed through application of fs-TA on films of the Pn derivatives on the semiconductor substrates. The proportion of triplet exciton contribution to the final optimised device photocurrent however was not calculated. For the pentacene dimer investigations a peak IQE of 127% was achieved on a $In_{60}Zn_{40}O$ substrate doped with a 0.1M Li+ solution, shown in Figure 12 [63]. This represents a clear verification of triplet contribution to photocurrent in this device, derived from intramolecular SF. However, due to the degree of band gap tuning required to accommodate low energy triplet excitons from the pentacene dimer, device efficiency was limited to 0.06% by the poor $V_{OC}$ of 0.22V.

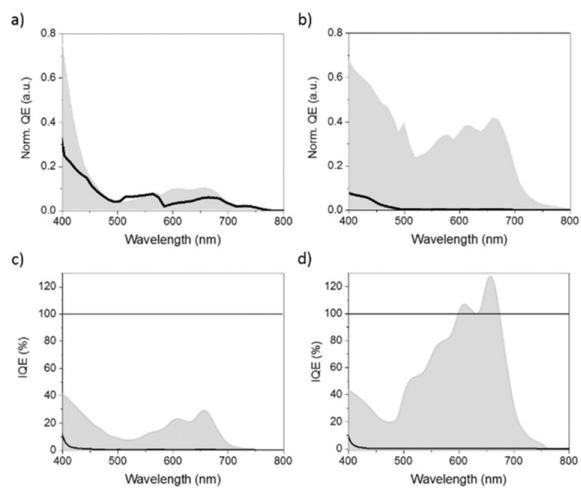

Figure 12: Top: Normalised EQE spectra for devices made with a pentacene derivative a) and pentacene dimer b). Bottom: IQE spectra for the pentacene derivative c) and dimer d). The solid black curve is for devices with 0.1M Li+ electrolyte and ZnO acceptor. Shaded grey regions are for devices with the same electrolyte and $In_{60}Zn_{40}O$ acceptor. Figure by Kunzmann *et al.*[63], Copyright (2021), reproduced with permission from Wiley Materials.

Since efforts to produce SF-enhanced DSSCs are in their early stages in comparison to OPVs and QD acceptor cells, it is difficult to determine their long-term viability at this stage. Due to the difficulty of producing accurate IQE measurements for this device type, there is a clear need for supplemental characterisation techniques such as MPC. Application of this technique to acene based devices as well as MFE characterisation of the DPIBF chromophore are recommended for reliable future measurements of triplet exciton contribution in these devices.

Verification of triplet sensitisation via conventional low-band gap acceptors, and of the viability of band gap tuning, are promising signs that practical devices based on the DSSC architecture may be feasible. The inability to accurately control the morphology of the chromophore layer due to the nanostructured acceptor layer presents a challenge to obtaining efficient intermolecular SF via monomer-based chromophores. Further investigation of dimers with larger triplet energies (i.e. tetracene-based dimers) may present a pathway toward efficient devices. The ability to tune triplet injection rates through adding spacer layers may also allow these endothermic chromophores to be viable despite slower SF rates than their exothermic counterparts.

### 3.2.4 Singlet Fission and Perovskite Acceptors

Thin-film perovskite-based PV cells have emerged in recent years as a potential competitor to silicon solar cells. Since their first iteration in 2009, they have rapidly increased in efficiency to a current record of 25.5%[95,96]. In addition to this high efficiency, perovskites display many of the potential advantages discussed earlier for QD based PV cells. These include tuneable band gaps, lower cost manufacturing processes (i.e., solution-based methods, Chemical Vapour Deposition), and compatibility with flexible substrates.

Poor device stability, and toxicity concerns raised by the presence of lead in most perovskites have thus-far limited commercial deployment of this technology. There are currently significant efforts to address these drawbacks in both academic and commercial sectors. Given the substantial improvements achieved to date, it is likely that perovskite solar cells will play a significant role in solar PV generation in future. Determining the suitability of SF-perovskite pairings in a device is therefore presents a worthwhile pathway for future research.

At the time of this review, an attempt to incorporate a SF donor and perovskite acceptor into a device has not been undertaken. A significant barrier to this approach is the high band gap energy of current champion efficiency perovskite absorbers $MAPbI_3$ (1.55-2.3eV)[97], and FAPbI3 (1.45-1.51eV)[98]. As of 2017, however perovskites alloyed perovskites of FAPbI3 and $CsSnI_3$ have achieved band gaps in the range of 1.24-1.51 eV[98], placing them within the triplet energy range of well-researched endothermic SF chromophores such as tetracene and DPIBF.

Whilst a SF device architecture utilising a perovskite material as the acceptor has not been realised to date, TA spectroscopy has been performed on TIPS-Pc/$MAPbI_3$ heterojunctions in two separate investigations by Lee *et al.* in 2017 and Guo *et al.* in 2020[60,61]. Both investigations displayed evidence of rapid SF within 1.1 ps. Rapid electron transfer from a product state to MAPI was reported to occur in both papers, with Guo *et al.* finding a transfer time of 1.5 ps. The nature of the product state involved was initially proposed to be the $T_1$ state by Lee *et al.*, but this was later disputed by Guo due to the endothermicity of this process (250meV). [60,99].. Instead, it was suggested that two-electron transfer from the correlated triplet pair state $^1(TT)$ was the primary mechanism responsible. The increase in electron density as a result of this process was estimated to be 38%. Hole transfer from the TIPS-Pc to MAPI occurred with a significantly slower time constant of 13.8 ns.

The results of these investigations, while not definitively proving the viability for SF-perovskite integration, do demonstrate efficient electron transfer between the organic/perovskite interface. The reverse process of triplet sensitisation of a SF-capable chromophore via photoexcited states in a 2D inorganic material has also been demonstrated in literature, both for $ReS_2$/Tetracene heterojunctions[100] and $MAFAPbI_3$/Rubrene heterojunctions[101]. Both systems are energetically suited for the TTA upconversion process rather than SF, with the latter demonstrating a TTA efficiency in excess of 3%[101]. Nevertheless, this work demonstrates that the interface between SF-chromophores and perovskites are capable of triplet transfer and thus may support efficient SF sensitisation with appropriate energy matching.

### 3.2.5 Singlet Fission sensitised Silicon Devices

Utilising silicon as an acceptor for SF presents one of the most promising prospects of realising a commercial SF boosted PV device. Since silicon based solar photovoltaic cells are currently the predominant technology in the solar energy market as of October 2021, further boosting silicon PV cell efficiency through SF may provide a significant efficiency benefit whilst also leveraging existing PV manufacturing infrastructure. The interface between silicon and the SF chromophore presents unique challenges however due to the presence of dangling silicon bonds that may act as recombination sites.

Hybrid a-Silicon/Quantum Dot Acceptor

Multiple different device structures have been trialled in literature in an attempt to resolve this issue, with the first by Ehrler *et al.* in 2012[102]. This structure utilises pentacene as the SF chromophore, with 1.1 eV PbSe nanocrystals used to harvest the triplet excitons. Electrons harvested from these excitons are then transferred from the nanocrystal layer to the amorphous silicon layer.

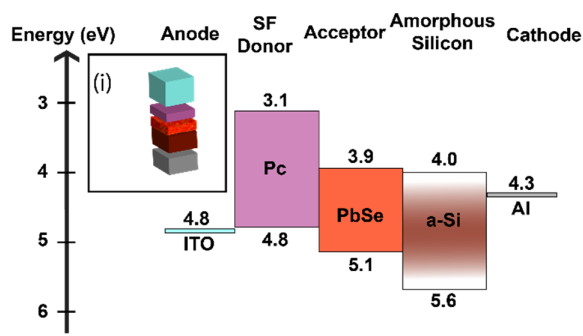

Figure 13: Hybrid a-Silicon/QD acceptor structure (left) and energy level diagram (right) for the device by Ehrler et al[102].

Amorphous silicon was used in place of the more efficient crystalline silicon due to its lower deposition cost and large absorption coefficient in thin-films. The nanocrystal layer was used as an intermediate between the pentacene and a-Si layer in an attempt to both provide a more ideal interface for exciton transfer and dissociation and also to protect the pentacene layer from the a-Si sputtering process. EQE modelling of the device revealed a photocurrent contribution from the pentacene layer, indicating that triplet harvesting from SF was occuring.

An improvement in EQE was also seen in this structure in the absorption range of a-Si compared to a control device with PbSe/a-Si absorption layers and no pentacene. This effect is attributed to improved hole extraction by the pentacene layer. The PbSe nanocrystals were also demonstrated to have some contribution to the device EQE through measurement of devices which lacked a pentacene layer, although the highest EQE was observed in the device with all three active layers. The measured efficiency of the device in this paper was low due to the use of undoped silicon, and the measured EQE of the device did not exceed 100% at any wavelength. It is likely however that the high $V_{loss}$ in the PbSe nanocrystals, known to limit the efficiency of QD solar cells, was also a contributing factor. Reproducing this result with an optimised device, and performing magnetic field dependent photocurrent measurements, would be useful in quantifying the benefit provided by SF.

Parallel Tandem Cell

The first attempt at integrating SF with crystalline silicon solar cells came in the form of a pentacene/Si parallel tandem device in 2017 by Pazos-Outón *et al.*[103]. This device architecture avoids complications from the silicon interface by indirectly coupling the SF chromophore to a c-Si solar cell via a parallel electrical connection. This architecture also provides some flexibility in band gap mismatch between the SF chromophore and low band gap absorber, since transfer of triplet excitons into the silicon is not required.

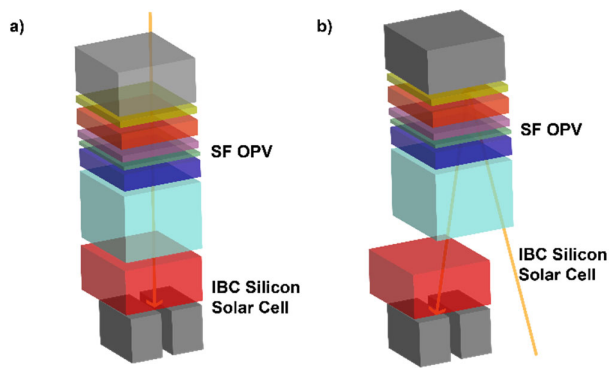

Figure 14: a) Parallel tandem cell with vertically stacked SF OPV top cell and Si bottom cell. b) Parallel tandem cell with angled reflective top contact SF OPV top cell and Si bottom cell. Device structure by Pazos-Outón *et al.*[103]

Device complexity and manufacturing cost will be increased however by this approach since at least one more electrical contact is needed. The device used in this paper consists of a standard pentacene/$C_{60}$ bilayer planar heterojunction structure as demonstrated by Congreve *et al.*, connected in parallel with a commercial interdigitated back contact (IBC) silicon solar cell. Two variants of the device were constructed, one with a semi-transparent pentacene top cell and another which utilised a reflective top electrode in the pentacene cell along with an angled light path to increase light absorption. EQE measurements of the transparent contact device clearly show a strong pentacene contribution. In the reflective device, EQE is demonstrated to exceed 100%, indicating a SF enhancement of device photocurrent. In the transparent device, EQE is limited by losses at the interfaces as well as by reflections from the semi-transparent top-electrode. The authors of this paper reported only a 50% transmittance through this top electrode, a value which will require significant improvement for a viable device to be realised. The authors also suggested that the number of electrodes could be reduced to three through utilising a silicon cell architecture that uses a conductive top electrode such as the HIT, a change which would reduce the complexity of the cell and may reduce parasitic voltage losses.

Direct Charge Transfer Architecture

The next approach taken in device implementation is that of direct exciton transfer from the SF chromophore to crystalline silicon. Due to the defined band gap of crystalline silicon as compared to amorphous silicon, triplet excitons must have at least 1.1 eV to be used in such a device, making pentacene unsuitable. For this reason, tetracene has been used in all directly SF sensitised c-Si devices.

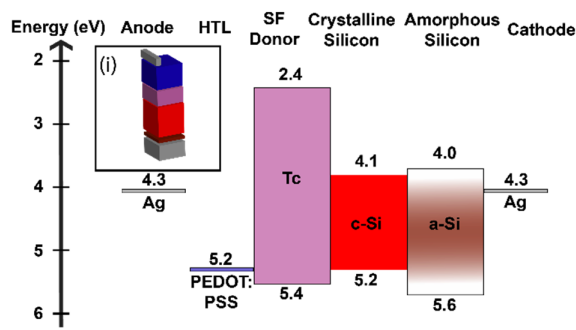

Figure 15: Tc/Si CT cell structure (inset) with energy level diagram. Device by Macqueen et al[104].

The first device to attempt this was produced by Macqueen *et al.* in 2018[104], which aimed to use charge transport as a mechanism for triplet transfer from tetracene into silicon. The structure consists of a n-doped c-Si layer bonded to an intrinsic a-Si layer as the low band gap absorber and electron acceptor. After etching to remove the $SiO_2$ layer, tetracene was directly deposited via thermal evaporation and a PEDOT:PSS layer was spin-coated onto the tetracene to aid hole extraction.

Efficiency measurements on this device indicated no net benefit from SF compared to the control device without tetracene (9.9% & 10% PCE respectively). EQE modelling of the devices suggested an exciton harvesting efficiency of 8% compared to a maximum of 200%. Whilst this suggests that triplet charge transfer does occur, it is not in sufficient quantities to provide a benefit in PV devices. A lack of driving force for triplet exciton dissociation was flagged as a potential reason for this, prompting alternative device structures that rely on triplet energy transfer.

Triplet Energy Transfer (Dexter process) Architecture

In 2019, a device fabricated by the Baldo Group achieved 133% exciton transfer efficiency utilising the Dexter energy transport mechanism in silicon[77]. In this type of device structure, it is not necessary for the SF chromophore to be electrically contacted since triplet excitons excite the silicon via a resonant energy transfer and do not dissociate into polarons. A hafnium oxynitride interlayer was used, both to passivate dangling silicon bonds and to act as a barrier to singlet energy transfer. Magnetic field dependent photocurrent and photoluminescence spectroscopy was used both to confirm the presence of SF and optimise the interlayer thickness.

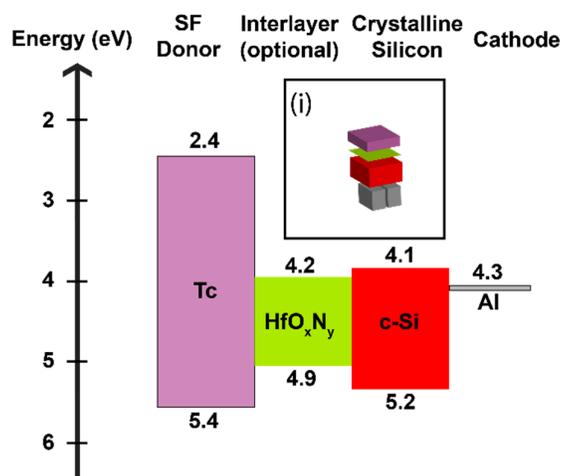

Figure 16: Tetracene/Silicon Triplet Energy Transfer architecture (left) and energy level diagram (right) for the device structure by Einzinger et al.[77].

The presence of a magnetic field effect in the cell showed a strong dependence on layer thickness, with decreasing MFE below 8 angstroms attributed to insufficient chemical passivation. Decreases above this critical thickness were attributed to the deactivation of triplet energy transfer, potentially due to deactivation of a tunnelling process. The exciton transfer efficiency of 133% at 30nm thickness is consistent with the peak value recorded in Tc OPVs of (127+/-18) % in a 25nm Tc layer by Wu et al. in 2014[81]. This suggests that the overall SF yield is limited by the slow rate of SF in tetracene in this case. The use of an inefficient silicon photodiode architecture likely also had a role in limiting device efficiency to 5% PCE. Further exploration of this architecture with a faster SF chromophore and more efficient silicon cell architecture would be useful in determining the viability of this approach for producing a commercial device.

The most recent attempt at a sensitised silicon device by Daiber et al.[105] used a similar approach, but omitted the hafnium oxynitride interlayer. The tetracene layer was instead deposited directly onto silicon which had been etched with HF to remove the SiOx layer. Under an $N_2$ atmosphere, the devices displayed magnetic field dependent photocurrent behaviour consistent with singlet exciton transfer. Upon aging under ambient laboratory conditions for a period of 5 days, evidence of triplet transfer efficiencies of ~36% (compared to 75% by Einzinger et al.) appeared in PL spectra. Encapsulated devices also displayed this behaviour after a longer period of time (~6 weeks), indicating that a change to either the tetracene film or the Si substrate upon oxygen interaction facilitates an increase in triplet transfer efficiency. This was an important result, which contrasts with a previous study which showed no evidence for triplet energy transfer from tetracene into silicon[106].

X-ray diffraction measurements on the Tc film indicate a change in morphology from the TCI polymorph to TCII. The TCII polymorph has been previously associated with a faster SF rate by Arias et al. due to a change in intermolecular coupling[87], although the effect on interface interactions has not been investigated in literature to date. It is possible that TCII also forms a more efficient triplet transfer interface with silicon than TCI although further investigation is needed to confirm this. The presence and effect of tetracene decay products, which will have formed upon oxygen exposure have also not been accounted for. It is not known thus far whether their role is only to alter the morphology of the unreacted tetracene or whether they directly influence the SF and triplet transfer process in this device. This behaviour is crucial in understanding both how to optimise such a device, and predicting the lifetime characteristics of a commercial device.

Given the apparent sensitivity of tetracene to morphology changes, it also cannot be ruled out that the exciton transfer efficiency gain observed by Einzinger *et al.*[77] was also related to a change in the Tc layer induced by the presence of the $HfO_xN_y$ interlayer. The independence of both approaches will need to be investigated in future research.

## 4. PROMISING PATHWAYS

In order to play a role in commercial PV devices, SF cell architectures must overcome a range of challenges. Development of chromophores with high SF yields and rapid SF rates, with a useful triplet energy level and sufficient chemical and photostability to survive long term usage is a necessary first step. Most device studies have utilized acenes due to their fulfillment of the first three of these criteria and the extensive body of published literature concerning their material and spectral properties. Due to their poor stability however, acenes are not likely to be the optimal candidate for a commercial SF device. It will be necessary to expand beyond acenes in future device studies to achieve this goal.

Acceptor materials and architectures must also enable efficient exciton transfer and dissociation, whilst negating loss mechanisms such as triplet charge annihilation or band gap mismatch. As discussed in section 3.2, not all candidate materials and architectures showed equal promise in meeting these criteria.

Organic acceptors have proved useful in characterising SF chromophores and will continue to do so in future. In planar structures, efficient triplet transfer has been demonstrated[38,81] but device efficiencies remain poor. In the more commercially popular bulk-heterojunction architecture, advantages provided by SF were further diminished by triplet charge annihilation. It therefore is unlikely at this stage that SF-OPVs will prove useful in commercial PV devices for power generation.

Both quantum dot and metal oxide semiconductor-based SF-sensitised solar cells also suffer from these issues, albeit to a lesser degree. $V_{OC}$ losses due to tail states in all disordered device structures including OPVs, DSSCs and QD based devices present significant obstacles at present to realising a high efficiency SF device with potential to breach the SQ limit.

Conversely, great potential has been demonstrated in recent years for the sensitisation of silicon with SF materials. Whilst efficiencies as of early 2021 do not exceed those of QD based SF devices, major limiting factors appear to be related to the unoptimised silicon acceptor rather than fundamental device limitations. Further optimisation of the acceptor structure, SF chromophore choice and donor acceptor interface may significantly improve device efficiencies and offers the possibility of exceeding current single junction thermodynamic efficiency limits. SF-sensitised silicon cells also offer the largest potential for commercialisation of all current approaches since conventional silicon solar cells occupy a dominant position in the commercial solar PV market as of 2021.

We described two approaches to sensitizing crystalline silicon with SF in a two-terminal device: charge transfer and exciton transfer. Photonically coupled downconverters are a third option which has garnered research interest[107,108], but no coupled crystalline silicon PV device has been reported to date. A photonically-coupled device requires making singlet fission bright by harvesting triplet states using either semiconductor or lanthanide-doped nanocrystals[25]. The photoluminescence quantum yield of these materials must exceed 50%, else all gains are lost. This is difficult. In the case of semiconductor nanocrystals, high quantum yields are usually achieved using thick shells which preclude triplet energy transfer. Separating the roles of triplet-harvesting and photon emission seems a likely necessity. Following photon emission, this energy must be harvested by the underlying silicon cell. Naïvely, one may expect only 50% of photons to be emitted in the right direction. However, total internal reflection and interaction with the evanescent wave of silicon may well conspire to provide efficient capture of photons[109]. The prospect remains challenging.

The charge-transfer strategy pioneered by MacQueen et al. requires the SF layer to play a dual role of hole conductor and exciton multiplier[104]. While promising, the device will necessarily suffer from triplet-charge annihilation, and charge carriers in the organic phase open up the device to further degradation pathways. This architecture has the further limitation of requiring a transparent top electrode.

The exciton-transfer approach pioneered by Einzinger et al. is the most promising[77]. In this approach, the SF layer has one job, but excitons must make their way to the interface with the underlying silicon. This interface must be simultaneously passivating, to allow the underlying rear-contacted cell to operate efficiently, and yet mediate exciton transfer. Ideally this transfer will occur in a concerted manner to minimize the number of free carriers in the organic phase. Some success was achieved with a thin hafnium oxynitride layer, but future, efficient devices will need a designed interface which mediates the exciton transfer from the world of molecularly-localized Frenkel excitons to the semiconductor world of weakly-bound, fleeting Wannier-Mott excitons.

SF-perovskite based cells are by far the least well explored of all of the acceptor materials discussed in this review. Nonetheless, investigations of SF-perovskite heterojunctions indicate that efficient triplet or charge transfer across the interface may be possible[110]. Furthermore, the rubrene MAPI interface has been found to improve the performance of perovskite solar cells, notwithstanding disagreements over the mechanism[111–113]. The advent of low band gap perovskite materials in recent years has also made energy matching with a SF acceptor significantly more viable. Given the significant future potential of perovskite PV cells in the commercial space, integration of a SF-chromophore into a perovskite cell also offers significant promise as a pathway towards commercialising the SF process in solar cells. As with silicon, triplet transfer across the interface with a bulk perovskite occurs as sequential charge transfer. In order to mediate concerted exciton transfer, one must engineer the interface to incorporate a degree of quantum confinement.

Whether silicon or perovskite, or charge or exciton transfer, the performance of a SF photovoltaic device will require precise engineering of the interface. Control of molecular excitons at a bulk semiconductor interface will likely impact on a range of technologies involving the interplay of electronics and photonics through matter. This is a rich vein which will be exploited in the coming years.

## ACKNOWLEDGMENTS

This work was supported by the Australian Centre for Advanced Photovoltaics and by the Australian Research Council Centre of Excellence in Exciton Science (funding grant no. CE170100026). M.I.C. acknowledges support from the Sydney Quantum Academy.

## AUTHOR DECLARATIONS

The authors have no conflicts to disclose.